\documentstyle[preprint,eqsecnum,aps]{revtex}
\tightenlines

\def\beq{\begin{equation}}
\def\eeq{\end{equation}}
\def\pa{\partial}

\def\la{\langle}
\def\ra{\rangle}

\def\pu{p^\mu}

\def\ku{k_\mu}
\def\varp{\varepsilon}
\def\bea{\arraycolsep .1em \begin{eqnarray}}
\def\eea{\end{eqnarray}}

\def\v#1{{\bf #1}}
\def\x#1{#1{(x,p)}}

\def\c#1{{\cal #1}}
\def\vp{{\bf p}}

\def\vk{{\bf k}}

\def\Tr{{\rm Tr}}

\def\[{\left [}
\def\]{\right ]}
\def\({\left (}
\def\){\right )}
\let\de=\delta
\let\eps=\epsilon
\let\no=\nonumber

\def\eq#1{(\ref{#1})}
\def\refr#1{$^{\cite{#1}}$}
\def\eqs#1{Eqs.(\ref{#1})}
\def\s0#1#2{\mbox{\small{$ \frac{#1}{#2} $}}}
\def\0#1#2{\frac{#1}{#2}}

\begin{document}
\title{\large {\bf Non-Abelian Medium Effects in Quark-Gluon Plasma}}
\author{Ji-sheng Chen$^{a,b},$
~~~~~~Jia-rong Li$^{b}$,~~~~~~Peng-fei Zhuang$^{a}$}
\address{$^a$Physics Department, Tsinghua University, Beijing 100084, P.R.
China.}
\address{$^b$Institute of Particle Physics,
Hua-Zhong Normal University,
Wuhan 430079, P.R.China.}
\maketitle
\date{June 15, 2001}

\begin{abstract}
~~~~Based on the kinetic theory, the non-Abelian medium property
of hot Quark-Gluon Plasma is investigated. The nonlinearity of the plasma
 comes from two aspects: The nonlinear
wave-wave interaction and self-interaction
of color field. The non-Abelian color permittivity is obtained by
expanding the kinetic equations to third
order. As an application, the nonlinear Landau
damping rate and the nonlinear eigenfrequency shift are calculated in the longwave
length limit.
\end{abstract}
\vskip 1.5cm
\section{Introduction}
Looking for Quark-Gluon Plasma (QGP) as an important aim of relativistic
heavy ion collisions
has inspired wide interest and stimulated extensive
studies\refr{Heinz,mro,Blaizot,Kelly,selikhov,bodeker,litim,arnold,Blaizot1}.
An urgent and important subject is to investigate medium effects in the hot
plasma. The high transverse momentum
particles and jets created by the hard scattering processes,
which propagate through the hot and dense matter produced by the soft
processes, can be used as
a clear and observable probe of QGP\refr{wangxn1,wangxn2}.
Theoretically, to study the behavior of high energy partons
through QGP one can construct a kind of response theory for QGP
to external current, which must be non-Abelian and hence
nonlinear\refr{koike}. The non-Abelian permittivity, as a characteristic quantity
of QGP medium, must be known.
\par
The kinetic theory is a powerful tool to investigate the medium
effects. The QGP kinetic theory which is fundamentally capable of
dealing with the non-equilibrium phenomena has been formulated
(see, for example, Ref.\cite{Heinz}),
and the Abelian-like permittivity of QGP in the linear
approximation has been given out\refr{Heinz}. However, in the linear
approximation, the property of QGP is quite similar to QED
plasma.
\par
The great difference between the QED and QGP plasma is that the former is
an Abelian medium while the latter is a non-Abelian one. It is not
sufficient to take only the Abelian dominance approximation or the
linear approximation in solving QGP kinetic equations. In order to
reflect the essential characteristic of QGP, one must go beyond
the linear approximation and take into account the
nonlinear wave-wave interaction and the
self-interaction of color field, as pointed out by
Ref.\cite{markov}.
\par
In this paper, we aim at discussing the nonlinear medium effects in QGP
based on the kinetic theory of chromodynamics. By using the
collisionless Vlasov kinetic equations, the induced color currents
up to third order are obtained. Then from the color field
equation including both the
nonlinear wave-wave interaction and the self-coupling term,
the color permittivity of QGP is calculated. As an application, the nonlinear
eigenfrequency shift and
the nonlinear damping rate are calculated for the mode $\v{k}=0$.
\par
The paper is organized in the following. In
Section \ref{one}, we review the kinetic
equations of QGP and discuss the expansion of the distribution function in
the color mean field. Section \ref{two} is devoted to obtaining the first
order, second order and third order color currents by solving
the kinetic equations. In Section \ref{three}, the general
expression of nonlinear permittivity
 is obtained by substituting the obtained color currents
into the color field equation.
The nonlinear Landau damping rate and
the nonlinear eigenfrequency shift are obtained from the color
permittivity in Section \ref{five}.
\par
The natural units and the Minkovski metric are used through the
paper, i.e., $\hbar =c=k_B=1$ and
$diag(g_{\mu\nu})=diag(g^{\mu\nu})=(1,-1,-1,-1)$. \vskip 1.5cm
\section{Kinetic Equations for fluctuations}\label{one}
\subsection{Mean Field and Kinetic Equations}
The color mean field equation in QGP is \bea\label{md} D_\mu
F^{\mu\nu }(x) =j^\nu (x), \eea with the color current \bea\label{cur}
 j^\nu
(x)=\0g2 \int \0{d^3p}{(2\pi )^3E}\[\(\x{Q_+}-\x{Q_-}\)+2it_a
f_{abc }G_{bc }(x,p)\] \eea
and the non-Abelian mean field stress tensor
 \bea
F^a_{\mu\nu }(x) =\pa _\mu A_\nu ^a (x) -\pa _\nu A^a _\mu (x) +g f_{abc } A^b
_\mu (x) A^c _\nu (x).
\eea
Here, $t^a,
a=1,2,\cdots, N_c^2-1,$ are the $SU(N_c)$ group generators in fundamental
representation and $f_{abc}$ are the corresponding structure constants.
\par
The quark and  anti-quark distribution functions $\x{Q_\pm}$ are
$N_c\times N_c$ matrices in color space, and behave as $\x{Q}\rightarrow
\x{U}\x{Q} U^\dagger (x,p)$ under the transformation operator $U(x,p)$. The
Gluon distribution function $\x{G}$ transforms as $\x{G}\rightarrow
\x{M}\x{G}M^\dagger (x,p)$ with $M_{ab}=\Tr [t^a \x{U} t^b U^\dagger
(x,p)]$.
\par
In the collisionless semi-classical limit with spin being neglected,
the distribution
functions $\x{Q_\pm}$ and $\x{G}$ satisfy the following transport
equations\refr{Heinz,mro}
\bea\label{Heinz}
\pu D_\mu \x{Q_\pm }\pm \0g2 p^\mu \0{\pa }{\pa p_\nu }\{F_{\mu\nu} (x),
\x{Q_\pm }\}=0,\no\\
\pu {\tilde D}_\mu \x{G}+ \0g2 p^\mu \0{\pa }{\pa p_\nu }\{{\cal F}_{\mu\nu}
(x),
\x{G}\}=0,
\eea
where $D_\mu $ and ${\tilde D}_\mu $ are the covariant
derivatives defined as
\bea
D_\mu =\pa _\mu -ig [A_\mu ,\cdots ],\no
\\
{\tilde D}_\mu =\pa _\mu -ig [\c{A}_\mu ,\cdots ].
\eea
The color mean field
$A_\mu (x)$ and $\c{A}_\mu$  and the field stress tensors $F_{\mu\nu } $ and
$\c{F}_{\mu\nu }$
have the following Lie-Algebra relations
\bea
&&A^\mu (x) =A^\mu _a(x) t^a,~~~~~ F^{\mu\nu }(x) =F^{\mu\nu }_a(x) t^a,
\no\\
&&\c{A}_{ab}(x) =-if_{abc}A_c ^\mu (x),~~~~~\c{F}^{\mu\nu
}_{ab}(x)=-if_{abc}F^{\mu\nu }_c (x).
\eea
\par
The \eqs{md} and (\ref{Heinz}) form a closed set of coupled equations.
It is easy to see  that the nonlinearity in QGP
results from two aspects. One is associated with the
nonlinear kinetic equations (\ref{Heinz}), and the other one is related to
the self-interaction term on the left hand side of mean field equation
(\ref{md}).
\par An usually used approximation to solve the above coupled equations is
the linearization.
Linearizing the left
hand side of mean field equation (\ref{Heinz}), i.e., neglecting the
self-interaction term,
was considered in Refs.\cite{zheng,zhang}, and linearizing the
kinetic equations
was discussed by Blaizot et al. in Ref.
\cite{Blaizot,chen1}.  What we will do in this paper is to discuss the
nonlinear effects in both the field and kinetic equations.
\par
\subsection{Equations of Fluctuations}
\par
For convenience to analyze the wave-wave interaction of fluctuations in QGP,
the mean fields and the distribution functions can be divided into regular
parts and fluctuation parts\refr{selikhov,litim,markov,zheng,tur}.
The quark, anti-quark and gluon distribution functions are decomposed as
\bea
&&\x{Q_\pm}=\x{{\bar Q_\pm}}+\x{q_\pm}
,
\no\\
&&\x{G}=\x{{\bar G}}+\x{g},
\eea
and in the meantime, the field $A$ and current $j^{\mu}$ are decomposed
as
\bea
A^a_\mu(x)={\bar A}^a_\mu (x) +A^{aT}_\mu(x), ~~~~~j^\mu _a (x) ={\bar
j}_\mu ^{a}(x)+j^{aT}_\mu (x).
\eea
Here the quantities having a bar represent the ensemble average of the
corresponding quantities,
and by definition the mean values of the fluctuations
vanish
\bea
&&\la q_{\pm}(x)\ra =0,~~~~~\la g(x)\ra=0,\no\\
&&\la A^T(x)\ra =0,
~~~~~ \la j^T(x)\ra=0.
\eea

\par
With the above decomposition, one can obtain the kinetic equations
for the fluctuations of distribution functions by subtracting the kinetic
equations for the regular distributions from the average of the kinetic
equations (\ref{Heinz}) $^{\cite{Heinz}}$
\bea\label{211}
\pu \pa _\mu && \x{q_\pm} =ig ([A^T_\mu (x),\x{q_\pm}]-\la
[A^T_\mu(x),\x{q_\pm}]\ra) \mp\012 g\pu \{F_{\mu\nu L}^T(x) ,\0 {\pa
\x{{\bar Q}_\pm}}{\pa p_\nu}\} \nonumber\\ &&\mp\012
g\pu(\{F_{\mu\nu L}^T(x),\0 {\pa \x{q_\pm }}{\pa p_\nu} \} -\la
\{F_{\mu\nu L}^T(x),\0 {\pa {q_\pm (x,p)}}{\pa p_\nu} \} \ra )
\nonumber\\ &&\pm \012 ig^2p^\mu (\{[A^T_\mu(x),A^T_\nu(x) ],\0
{\pa {q_\pm (x,p)}}{\pa p_\nu}\} -\la \{[A^T_\mu(x),A^T_\nu(x) ]
,\0 {\pa {q_\pm (x,p)}}{\pa p_\nu}\}\ra ) \nonumber\\ && \pm \012
ig^2p^\mu \{[A^T_\mu(x),A^T_\nu(x) ] -\la [A^T_\mu(x),A^T_\nu(x) ]
\ra , \0 {\pa \x{{\bar Q}_\pm }}{\pa p_\nu}\},
\no\\
\pu \pa _\mu && \x{g} =ig ([{\cal A}^T_\mu(x),\x{g}]-\la [{\cal
A}^T_\mu(x),\x{g}]\ra) -\012 g\pu \{{\cal F}_{\mu\nu L}^T(x) ,\0
{\pa \x{{\bar G}}}{\pa p_\nu}\} \nonumber\\ &&-\012 g\pu(\{{\cal
F}_{\mu\nu L}^T(x),\0 {\pa \x{g}}{\pa p_\nu} \} -\la \{{\cal
F}_{\mu\nu L}^T(x),\0 {\pa {g(x,p)}}{\pa p_\nu} \} \ra )
\nonumber\\ &&+\012 ig^2p^\mu (\{[{\cal A}^T_\mu(x),{\cal
A}^T_\nu(x) ],\0 {\pa {g(x,p)}}{\pa p_\nu}\} -\la \{[{\cal
A}^T_\mu(x),{\cal A}^T_\nu(x) ] ,\0 {\pa {g(x,p)}}{\pa p_\nu}\}\ra
) \nonumber\\ && +\012 ig^2p^\mu \{[{\cal A}^T_\mu(x),{\cal
A}^T_\nu(x) ] -\la [{\cal A}^T_\mu(x),{\cal A}^T_\nu(x) ] \ra , \0
{\pa \x{{\bar G}}}{\pa p_\nu}\}, \eea
where the index $L$ denote the linear term of $F_{\mu
\nu}$ in $A^T_\mu$.
\par
The fluctuation part of the color field  satisfies the following
Yang-Mills equation \bea\label{yang} \pa _\mu
&&F^{T\mu\nu}_L+j^{T(1)\nu} =- j^{T\nu}_{NL} +ig \pa _\mu
([A^{T\mu},A^{T\nu}]-\la [A^{T\mu},A^{T\nu}]\ra )\nonumber\\&& +ig
([A^T_\mu , F^{T\mu \nu }_L ]-\la [A^T_\mu , F^{T\mu \nu}_L ] \ra)
+g^2([A^T_\mu ,[A^{T\mu},A^{T\nu} ]]-\la [A^T_\mu
,[A^{T\mu},A^{T\nu}]] \ra ), \eea where  the current fluctuation $j^{T\mu }$
is also divided into a linear term $j^{T(1)\nu }$ and a nonlinear term
$j^{T\nu}_{NL}$. By expanding the
fluctuation of the distribution functions in terms of the fluctuations of the
color field $A^{aT}_\mu$
\label{expansion}
\bea
\x{q_\pm}=\sum _{n=1}^{\infty}q^{(n)}_\pm(x,p),
\no\\
\x{g}=\sum _{n=1}^{\infty}g^{(n)}(x,p),
\eea
the nonlinear color current can be expressed as
\bea j^{T\mu}=\sum _{n=2}^{\infty}j^{T(n)\mu}, \eea
with
\bea\label{curo} j^{T(n)\mu}(x)=gt^a\int\0 {d^3p}{(2\pi
)^3p^0}p^\mu \[{\em tr} \(t^a\left
\{q_+^{(n)}(x,p)-q_-^{(n)}(x,p)\right \}\)+{\em Tr}
(T^ag^{(n)}(x,p))\], \no\\
n=2,3,4,\cdots.
\eea
\par
Substituting the above expansion of fluctuations in $A^T_\mu $ into the
kinetic equations (\ref{211}), one obtains the hierarchies of kinetic
equations for the distribution functions $q_\pm ^{(n)}$ and $g ^{(n)}$.
\par
By using the regular distributions and regular field as input, which can be
calculated from the corresponding kinetic equations and mean field equation,
one can in principle solve the above kinetic and mean field equations for
the fluctuations\refr{selikhov,litim,tur,landau}. For simplification, we
will consider in the following the fluctuations above the colorless
equilibrium state. We take ${\bar A}=0$ and
Fermi-Dirac or Bose-Einstein distribution
function for ${\bar Q}_\pm$ and ${\bar G}$
\bea {\bar Q}_{\pm}=\0{2N_f}{e^{\beta
p\cdot u}+1},~~~{\bar G}=\0{2}{e^{\beta p\cdot u}-1},\eea where
$u=(1,0,0,0)$ is the local four-velocity of the plasma, $N_f$ is
the the number of flavors for quarks.
Since we are interested in the fluctuations only, the index $T$ for the field and
current will be omitted
from now on.
\par
It's convenient to discuss the fluctuations in momentum space
 through the Fourier transformation,  \bea
q_\pm (x,p)=\int \0 {d^4 k}{(2\pi )^4}q_\pm (k,p) e^{-ikx}.\eea
We choose temporary axis gauge, i.e., $A_0=0$,  and
 consider only the longitudinal field excitation,
$A^i(k)=\0{k^iA(k)}{K}$ with $K=|\vk |$.
\vskip 1.5cm
\section{Linear and nonlinear color currents}\label{two}
\subsection{Linear Approximation}
\par
In order to investigate the nonlinear property of QGP, one should solve the
kinetic hierarchies for the distributions $q_\pm ^{(n)}$ and $g^{(n)}$ order
by order. Let's first review the linear approximation which
 has been discussed in many previous
works. The dispersion relation obtained here will be
used to perform the relevant integrals
in the end of this paper.

>From the kinetic equations for the first order fluctuations of the distribution
functions, it is easy to write down
\bea\label{twoone}
q_\pm^{(1)}(k,p)&&=\mp\0 {k^0p^i A^i (k)}{p\cdot k+ip^0\epsilon ^+}\cdot \0 {\pa
{\bar Q}_\pm(p^0)}{\pa p^0}
,\no\\
g^{(1)}(k,p)&&=-\0 {k^0p^i {\cal A}^i (k)}{p\cdot k+ip^0\epsilon ^+}\cdot
\0{\pa {\bar G(p^0)}}{\pa p^0}
.
\eea
Substituting these first order fluctuations into \eq{curo}, one obtains
 the first order color current
by noting $\Tr T^aT^b=N_c\de ^{ab}$.
Then from the mean field equation to the first order,
one has the nontrivial solution
\bea\label{firs}
-\omega ^2\varp _L(\omega, {\bf k}) A(k)=0,
\eea
with the well-known linear dispersion relation
\bea\label{longdis}
\varp _L(\omega, {\bf k})=1+\0{3\omega ^2_p}{K^2}[1-\0{\omega }{2K}(\ln
|\0{K+\omega }{K-\omega }|-i\pi \theta (K-\omega
))]=0,
\eea
where
 \bea \omega _p=\sqrt{\0{(N_f+2N_c)}{18}g^2T^2}\eea is the plasma frequency.
In the longwave length region, the dispersion relation
is reduced to
\bea\label{longw}\omega ^2=\omega
^2_p+\035 k^2. \eea
It is necessary to note that the linear dispersion relation (\ref{longdis})
 means that the eigenwaves in QGP are always timelike, they can not exchange
energy directly with the particles
in the plasma. Therefore in the linear approximation, the eigenwaves in QGP do not
undergo damping\refr{Heinz}.
\par
To obtain the nonlinear effects in the medium, one should solve the kinetic
equations to
nonlinear orders, i.e., to go beyond the linear order approximation.
\subsection{Second Order and Third Order Color Currents}
In this subsection, the second order and third order color
currents will be calculated. The following shorthands \bea \sum
_{k=k_1+k_2}&&=\int \de (k-k_1-k_2) dk_1dk_2\0{1}{(2\pi)^4},\\
\sum _{k=k_1+k_2+k_3}&&=\int \de (k-k_1-k_2-k_3)
dk_1dk_2dk_3\0{1}{(2\pi)^8}, \eea are used for notational
simplification. \par The second order kinetic equation for the quark
distribution is the following \bea\label{secdis} -ip^\mu k_\mu
&&q^{(2)}_+(k,p) =igp^i\sum _{k_1+k_2=k}
\([A_i(k_1),q_+^{(1)}(k_2,p)]-\la [A_i(k_1,p),q_+^{(1)}(k_2,p)]\ra
\)\nonumber\\ && + \012 gp^\mu \sum _{k_1+k_2=k} \( k_{1\mu}
\{A_i(k_1), \0{\pa q^{(1)}_{+}(k_2,p)}{\pa p_i}\}- \la k _{1\mu}
\{A_i(k_1), \0{\pa q^{(1)}_{+}(k_2,p)}{\pa p_i} \}\ra \)
\nonumber\\ &&- \0i2 gp^i\sum _{k=k_1+k_2}\(\{k_{1\nu} A_i(k_1),
\0{\pa q^{(1)}_{+ }(k_2,p)}{\pa p_\nu}\}-\la \{k_{1\nu} A_i(k_1),
\0{\pa q^{(1)}_{+ }(k_2,p)}{\pa p_\nu}\}\ra \)\no\\ && +
\0i2gp^i\sum_{k=k_1+k_2+k_3}\(\{[A_i(k_1),A_j(k_2)],\0{\pa {\bar
Q}_{+}(k_3,p)}{\pa p_j}\}-\la \{[A_i(k_1),A_j(k_2)],\0{\pa {\bar
Q}_{+ }(k_3,p)}{\pa p_j}\}\ra \). \eea  The kinetic
equations for the second order fluctuations of the anti-quark
 and gluon distribution functions
can be written out similarly to \eq{secdis} with opposite
 signs for the terms related to $\{\cdots \}$ for anti-quarks and with the
replacement of $q, A$ and
$F_{\mu\nu }$ by $G, {\cal A}$ and ${\cal F}_{\mu\nu }$ for gluons.

\par
Inserting the second order distribution functions into Eq.\eq{curo},
 one obtains the second order color current
\bea\label{sec} &&j^{T(2)ka}=\nonumber\\ &&-ig^3f_{abc}\sum
_{k=k_1+k_2}\int \0{d^3p}{(2\pi )^3p^0}\0{p^ip^jp^k}{p\cdot
k+ip^0\varp}\cdot \0{\omega _2\pa _{p_0}{\cal D}_{eq}} {p\cdot
k_2+ip^0\varp }\left (A_i^b(k_1)A_j^c(k_2)-\la
A_i^b(k_1)A_j^c(k_2)\ra \right ), \eea with the
effective equilibrium distribution function
\bea {\cal
D}_{eq}=\012N_f({\bar Q}_++{\bar Q}_- )+N_c{\bar G}.\eea
\par
The third order color current can be obtained similarly
\bea\label{thcu}
&&j^{T(3)la}(k)=\nonumber\\ &&\sum _{k=k_1+k_2+k_3} \Sigma
_{kk_1k_2k_3}^{abdeijkl}\(A_i^b(k_3)A_j^d(k_1)A^e_k(k_2)-A_i^b(k_3)\la
A_j^d(k_1)A^e_k(k_2)\ra -\la A_i^b(k_3)A_j^d(k_1)A^e_k(k_2)\ra \),
\eea with
\bea
\Sigma _{kk_1k_2k_3}^{abdeijkl}=
&&f^{abc}f^{cde}\Sigma_{kk_1k_2k_3}^{(I)ijkl}+\de
^{ab}\de ^{de}\Sigma
_{kk_1k_2k_3}^{(II)ijkl}+d^{abc}d^{cde}\Sigma
_{kk_1k_2k_3}^{(III)ijkl}\nonumber\\ &&+(\de ^{ab}\de ^{de}+\de
^{ad}\de^ {be}+\de ^{ae}\de ^{db})\Sigma
_{kk_1k_2k_3}^{(VI)ijkl},\no\\ \Sigma
_{kk_1k_2k_3}^{(I)ijkl}=&&-g^4\int \0{d^3p}{(2\pi )^3p^0}
\0{p^ip^jp^kp^l}{p\cdot k +ip^0\epsilon }\0{1}{p\cdot
(k_1+k_2)+ip^0\epsilon }\0{\omega _2\pa _{p_0}{\cal D}_{eq}}{p\cdot
k_2 +ip^0\epsilon} ,\no\\ \Sigma
_{kk_1k_2k_3}^{(II)ijkl}=&&\0{2}{N_c}\Sigma
_{kk_1k_2k_3}^{(III)ijkl}\nonumber\\=&&-\0{g^4}{2N_c}\int
\0{d^3p}{(2\pi )^3p^0}\0{p^l\chi ^{i\mu }(k_3,p)}{p\cdot k
+ip^0\epsilon}\0{\pa }{\pa p^\mu }\left (\0{\chi ^{j\nu
}(k_1,p)}{p\cdot (k_1+k_2)+ip^0\epsilon }\0{\pa }{\pa p^\nu
}(\0{\omega _2\pa _p^0{\cal D}_{eq}}{p\cdot k_2
+ip^0\epsilon})\right ),\no\\ \Sigma
_{kk_1k_2k_3}^{(VI)ijkl}=&&-\0{g^4}{2}\int \0{d^3p}{(2\pi
)^3p^0} \0{p^l\chi ^{i\mu }(k_3,p)}{p\cdot k +ip^0\epsilon}\0{\pa
}{\pa p^\mu } \left (\0{\chi ^{j\nu }(k_1,p)}{p\cdot
(k_1+k_2)+ip^0\epsilon }\0{\pa }{\pa p^\nu }(\0{\omega _2\pa
_p^0{\bar G}}{p\cdot k_2 +ip^0\epsilon})\right ), \eea
and \bea \chi^{\mu \nu }(k,p)=(p\cdot k)g^{\mu
\nu}-p^\mu k^\nu, \eea
where $d_{bce}$ are the
$SU(N_c)$ symmetric structure constants. From \eqs{sec} and \eq{thcu}, one can
see that the second and third order color currents come from the
response of particles to the secondary waves resulting from the
nonlinear interaction of eigenwaves.
\par
Before making further calculations, one should note that on the
right hand side of \eq{thcu}, the term with
$\Sigma_{kk_1k_2k_3}^{(I)ijkl}$ is purely non-Abelian, while the
others are Abelian terms. Since  the soft excitation
momentum $k\sim gT$ and the particle momentum $p\sim T$\refr{Blaizot,bodeker},
one
can easily conclude that the leading non-Abelian term
is proportional to $g^2$, while the Abelian terms
 are proportional to the higher orders of the coupling constant $g$: $\Sigma
_{kk_1k_2k_3}^{(II)ijkl}\sim \Sigma
_{kk_1k_2k_3}^{(III)ijkl}\sim g^3$ and $g^4$ and $\Sigma
_{kk_1k_2k_3}^{(VI)ijkl}\sim g^3$. \vskip 1.5cm
\section{nonlinear permittivity}\label{three}
Now we turn to calculate the non-Abelian permittivity from the
mean field equation to the third order
\bea\label{thirdf} -\ku k^\mu
A^{i}(k)&&+k_jk^iA^{j}(k)+j^{T(1)i}(k)=\nonumber\\g^2\sum
_{k=k_1+k_2+k_3}&&\left ([A_j(k_1),[A^{j}(k_2),A^{i}(k_3)]]-\la
[A_j(k_1),[A^{j}(k_2),A^{i}(k_3)]]\ra \right )-j^{Ti}_{NL}(k),\eea
with \bea j^{Ti}_{NL}(k)=j^{T(2)i}(k)+j^{T(3)i}(k). \eea
\par
>From \eq{thirdf}, one can see that the left hand side  is
Abelian-like, while the right hand side is non-Abelian which
results from two different origins. The first term on the right
hand side of \eq{thirdf} with the summation represents the color
self-interaction.
Some of our previous works \cite{zheng,zhang} neglected this nonlinearity.
The second non-Abelian property in \eq{thirdf} is from the nonlinear current
$j^{Ti}_\mu $. This nonlinear wave-wave interaction is not taken into
account in Refs. \cite{Blaizot,Kelly,chen}.
 Here we will discuss
the solution of the field
equation by considering the color self-interaction and the nonlinear
currents simultaneously.
\par
Substituting $j^{T(2)}(k)$ and $j^{T(3)}(k)$ into
the field equation \eq{thirdf}, one gets the relation
\bea\label{third}&& -\omega ^2 A(k) +{\bf j}^{T(1)}\cdot \0{\bf
k}{K}=\no\\&& ig^3t^af^{abc}\sum _{k=k_1+k_2}\int \0{d^3p}{(2\pi
)^3p^0}\0{(\v{p}\cdot
\v{k})}{K}\0{(\v{p}\cdot\v{k_1})}{K_1}\0{(\v{p}\cdot\v{k_2})}{K_2}\0{1}{p\cdot
k+ip^00^+}\0{\omega _2\0{\pa {\cal D}_{eq}}{\pa p_0}}{p\cdot k_2
+ip^00^+}\no\\&&~~~~~ \times\( A^b(k_1)A^c(k_2)-\la
A^b(k_1)A^c(k_2)\ra \)\no\\ &&-g^2t^a\sum _{k=k_1+k_2+k_3}\left
[g^2\int \0{d^3p}{(2\pi
)^3p^0}\0{(\v{p}\cdot\v{k_1})(\v{p}\cdot\v{k_2})(\v{p}\cdot\v{k_3})({\bf
p}\cdot {\bf k})}{K_1K_2K_3K} \0{1}{p\cdot k
+ip^0\epsilon}\right.\no\\&&~~~~~~~~~~~\times\left. \0{1}{p\cdot
(k_1+k_2)+ip^0\epsilon }\0{\omega _2\pa _{p_0}{\cal D}_{eq}}{p\cdot
k_2 +ip^0\epsilon}+\0{({\bf k}_1\cdot {\bf k}_2)}{K_1K_2}\0{({\bf
k}\cdot{\bf
k}_3)}{KK_3}\right ]
 \nonumber\\
&& ~~~~~\times f^{abc}f^{cde} \left
(A^b(k_3)A^d(k_1)A^e(k_2)-A^b(k_3)\la A^d(k_1)A^e(k_2) \ra-\la
A^b(k_3)A^d(k_1)A^e(k_2)\ra \right ). \eea
 Obviously the first and second integrals reflect the
three-wave process ($k\rightleftharpoons k_1+k_2$) and the four-wave
 ($k\rightleftharpoons
k_1+k_2+k_3$) process, respectively.
\par
It is well known that it is difficult to solve the general Yang-Mills equation
 because of the color field self-coupling term.
However,
 if the phase of excitations in QGP
is random (Gauss-like distributed), the correlation of even random
quantities can be approximately expressed as the product of
correlations of two quantities\refr{tur,stinko}, and the solution
of the field equation can be greatly simplified. Concretely, after
multiplying both sides of \eq{third} with $A^g(k')$, taking
average with respect to the Gauss-like random phase, and then using the
following relations\refr{stinko}
\bea
I^{dg} _k = \la A^dA^g\ra _{\omega , {\bf k}},~~~~~&& \la
A^d(k)A^g(k')\ra=(2\pi)^4I^{dg}_k \de
(k+k'),
\\
\la A^a(k_1)A^b(k_2) A^c(k_3)A^d(k_4)\ra =&&(2\pi)^8\de
(k_1+k_2+k_3+k_4)\times \(\de (k_1+k_2)I^{ab}_{k_1}I^{cd}_{k_3}
\right.\nonumber\\
&&\left. +\de
(k_1+k_3)I^{ac}_{k_1}I^{bd}_{k_2}+\de
(k_1+k_4)I^{ad}_{k_1}I^{bc}_{k_2}\), \eea
the equation \eq{third} is reduced to \bea\label{nper}
\left (\varp _L \de _{ad}+ \varp ^{\ast }_{ad} \right ) I^{dg}_k
=0, \eea where $\varp _L$ is the linear permittivity given in
\eq{longdis}, and the nonlinear permittivity $\varp ^{\ast
}_{ad} $ contains two parts
\label{nperr}
\bea\label{nper3}
\varp ^{\ast }_{ad}= \varp ^{R\ast }_{ad}+\varp
^{S
\ast }_{ad}\no,
\eea
with
\bea
\varp ^{S\ast }_{ad}=&&-\0{g^2}{\omega ^2}\int\left [ f^{abc} f^{ced}I^{be}
_{k_1}
+ \left (\0{\vk
\cdot \vk _1}{KK_1}\right )^2 (f^{abc}f^{cde}I^{be}_{k_1} +
f^{adc}f^{cbe}I^{be}_{k_1} )\right ]\0{dk_1}{(2\pi )^4},
\no\\\label{nper5}
\varp ^{R\ast }_{ad}=&& \0{g^4}{\omega ^2}\int\0{d^3p\pa _{p_0}
{\cal D}_{eq}}{(2\pi )^3p^0} f^{abc}\left ( \0{ \vp\cdot \vk _1
}{K_1} \right )^2\left ( \0{ \vp\cdot \vk }{K} \right
)^2\0{1}{p\cdot k+ip^0\eps } \nonumber\\ &&\times\0{1}{p\cdot
(k-k_1)+ip^0\eps }\left (\0{\omega }{p\cdot k+ip^0\eps
}f^{ced}I^{be}_{k_1}+\0{\omega _1}{p\cdot k_1+ip^0\eps
}f^{cde}I^{be}_{k_1} \right ) \0{dk_1}{(2\pi )^4}.
\eea
Here $\varp ^{S\ast }_{ad}$ and $\varp
^{R\ast }_{ad}$ correspond to the first and second term on
the right hand side of \eq{thirdf}, respectively. They reflect the  purely
non-Abelian medium effects
 of QGP.  From Eq. \eq{nper5},
the nonlinear permittivity $\varp ^{\ast }_{ad}$ is a matrix in
color space because of the non-Abelian
$SU(N_c)$ structure constants. Its value depends on the correlation
strength $I^{bc}_k=\la A^bA^c \ra _k $ and the
 dispersion relation $\omega (\vk ) $ of the collective
excitation.
\par
For simplification, we consider only the correlation between the same colors
and take a diagonal matrix in the color space
\bea
\la A^bA^c\ra _k=-\0 {\pi }{\omega ^2}(\de (\omega -\omega _\vk
)+\de (\omega +\omega _\vk ))I_\vk \de _{bc}, \eea where $I_\vk$
is the intensity of the correlation
with frequency $\omega _\vk $.
For fluctuations at thermal level, one can take $I_\vk=4\pi
T$\refr{markov,zhang,stinko}.
Then the nonlinear contributions $\varp ^{S\ast }_{ad}$ and $\varp
^{R\ast }_{ad}$ to the color permittivity are further reduced to
\bea\label{fa}
\varp ^{S\ast }=&&-\0{2g^2N_c}{\omega ^2(2\pi )^4}\int \left(1- (\0{\vk \cdot \vk
_1 }{KK_1})^2 \right )\0{\pi I_{\vk_1}}{\omega
_{\vk _1}^2}d^3k_1;\no\\\label{ast}
\varp ^{R\ast }=&&\0{2g^4 N_c }{\omega ^2} \int \0{d^3p\pa ^0
_p {\cal D}_{eq} }{(2\pi )^7p^0}
 \left (\0{\vp \cdot \vk _1}{K_1} \right )^2\left (\0{\vp \cdot \vk }{K}
\right )^2 \0{1 }{p\cdot k+ip^0 \eps } \nonumber\\&&\times \left(
\0{1}{p\cdot (k-k_1)+ip^0\eps }\0{\omega _0 }{p\cdot k+ip^0 \eps
}-\0{1}{p\cdot (k+k_1)+ip^0\eps } \0{\omega _{\vk _1}}{p\cdot
k_1+ip^0 \eps }\right )
 \0 {\pi I_{\vk_1} }{\omega ^2_{\vk _1}}d^3k_1.
\eea
The four factors $\0{1}{p\cdot k
+ip^0\eps }$, $\0{1}{p\cdot k _1 +ip^0\eps }$,$\0{1}{p\cdot (k+k_1)+ip^0\eps
}$, $\0{1}{p\cdot (k-k_1)
+ip^0\eps }$ in the expression of $\varp ^{R\ast }$ have different contributions
to
the integral. The first three factors have only real contributions
because the eigenwaves are always timelike,
 while the last one has
both real and imaginary contributions to $\varp ^R$ due to the well-know
formula
\bea\label{crucial}
\0{1}{p\cdot (k-k_1) +ip^0\eps }=P\0{1}{p\cdot (k-k_1)}-i\0{\pi}{p^0}
\de (\omega _\vk - \omega _{\vk _1}-{\bf v}
 \cdot (\vk -\vk _1)),\eea
where ${\bf v}=\vp /p^0$.
The imaginary part of $\varp ^{R\ast }$ indicates that the secondary waves
resulting from the nonlinear interactions can exchange energy with the plasma
particles, which corresponds to the
nonlinear Landau damping.
\par
To perform the above integrals over $\vk_1$ and $\v{p}$,
we can choose spherical polar coordinates and take
the polar axis of $\v{k}_1$ in the direction of $\v{k}$ and the polar axis
of $\v{p}$ in the direction of $\v{k}_1$.  Considering the soft excitation
in QGP\refr{bodeker}, the up and low limits of $K_1$
can be chosen to be $gT$ and $g^2T$, respectively.

By using Eq. \eq{crucial} and the integral  \bea
-\int _0^\infty p_0^2\pa _{p_0}{\cal D}_{eq}dp_0=\0{(N_f+2N_c)\pi
^2 T^2}{3},
\eea
the nonlinear color permittivities $\varp ^{S\ast }$ and $\varp ^{R\ast }$
 are finally written as one dimensional integrals
\bea \label{1sp} \varp ^{S\ast }=&&-\0{64\pi ^3g^2N_c }{3(2\pi
)^4\omega _p^2}\int ^{gT}_{g^2T} \0{1}{\omega ^2_{\vk
_1}}K_1^2dK_1 ,\no\\\label{2rp} Re(\varp ^{R\ast
})=&&\nonumber\\-&&\0{8g^4N_c\pi ^3 T^3(N_f+2N_c) }{9(2\pi
)^4\omega ^3 _p}\left [ \int _{g^2T}^{gT}\0{K_1^2dK_1}{\omega
^2_{\vk _1}} \0{2(\omega _{\vk _1}-\omega _p)}{K^2_1}\left
(1-\0{\omega _{\vk _1}-\omega _p}{2K_1}\ln \left |\0{\omega
_p-\omega _{\vk _1}-K_1}{\omega _p-\omega _{\vk _1}+K_1}\right
|\right )\right .\nonumber\\ &&\left.-\int ^{gT}_{g^2T}
\0{K_1^2dK_1}{\omega _{\vk _1}}\0{2}{K_1^2}\left (1+\0{\omega
^2_{\vk _1}}{2\omega _pK_1}\ln \left |\0{\omega _{\vk
_1}-K_1}{{\omega _{\vk _1}+K_1}}\right |-\0{(\omega _{\vk
_1}-\omega _p)^2}{2\omega _pK_1}\ln \left |\0{\omega _{\vk
_1}-\omega _p-K_1}{\omega _{\vk _1}-\omega _p+K_1}\right | \right
)\right ] ,
\no\\
Im(\varp ^{R\ast })=&&\0{8g^4N_c\pi ^4 T^3(N_f+2N_c) }{9(2\pi )^4\omega _p^4}\int
^{gT}_{g^2T}
\0{1}{K_1}dK_1 (\omega _{\vk _1}-\omega _p)^2\0{1}{\omega _{\vk _1}}.
\eea
\par
The contributions of $\varp ^{R\ast }$ and $\varp ^{S\ast }$ can be seen
quantitatively by
making a rough numerical estimate.
Using the dispersion relation \eq{longw}
in the longwave length limit, the total nonlinear permittivity is
\bea\label{final}
\varp ^{\ast }=&&\varp ^{S\ast }+Re(\varp ^{R\ast })+{\em i}Im(\varp ^{R\ast
})
\approx
-0.98g-1.91g+0.61g{\em i}=-2.89g+0.61g{\em i}.
\eea

\vskip 1.5cm
\section{conclusion}\label{five}

\par
Based on the chromodynamic kinetic and mean field equations, the nonlinearity in
QGP
has been analyzed.
One nonlinearity comes from the response of
particles to the secondary waves resulting from the nonlinear
interaction, and the other nonlinearity is due to the non-Abelian
self-interaction in mean field equation.
The non-Abelian permittivity due to the
two nonlinearities has been calculated by considering the second and third
order color currents.
>From its general expression,
it depends on the dispersion modes $\omega _\vk$
and the correlation strength of color field. The non-Abelian
property is reflected in the presence of the $SU(N_c)$ structure constants
$f_{abc}$.
With the given color permittivity, it is possible to
 further analyze the non-Abelian response of QGP to external
currents\refr{koike}.

\par
As an application of the nonlinear color permittivity \eq{final},
we can obtain the nonlinear
Landau damping rate\refr{tur}
\bea\label{re}
\gamma =
-\0{Im (\varp ^{\ast })}{\0{\pa \varp _L}{\pa \omega }|_{\omega
=\omega _\vk }}=0.17g^2T,
\eea
and the nonlinear eigenfrequency shift
\bea
\Delta \omega =
-\0{Re (\varp ^{\ast })}{\0{\pa \varp _L}{\pa \omega }|_{\omega
=\omega _\vk }}=0.83g^2T,\eea
which
characterizes the nonlinear effect in QGP too.
\acknowledgments{The work was supported in part by the NSFC under Grant No.
19925519, 10135030 and 10175026.}

\end{document}